\documentclass[
reprint,
preprintnumbers,
nofootinbib,
amsmath,amssymb,
aps
]{revtex4-2}

\usepackage{graphicx}
\usepackage{amsfonts}
\usepackage{accents}
\usepackage{color}
\usepackage[dvipsnames]{xcolor}
\usepackage{dcolumn}
\usepackage{bm}
\usepackage{hyperref}

\definecolor{RbCs}{HTML}{800021}
\definecolor{HSi}{HTML}{ff0000}
\definecolor{SrCs}{HTML}{8c0029}
\definecolor{YbSr}{HTML}{376631}
\definecolor{SrSi}{HTML}{00a86b}
\definecolor{YbE3E2}{HTML}{228b22}
\definecolor{YbE3Sr}{HTML}{66991a}
\definecolor{AlHg}{HTML}{004d00}
\definecolor{auriga}{HTML}{006cb3}
\definecolor{sne}{HTML}{ff7808}
\definecolor{qcd}{HTML}{b6cf1a}
\definecolor{earth}{HTML}{00c4d1}
\definecolor{edm}{HTML}{a15f59}
\definecolor{comag}{HTML}{ee9ebe}
\definecolor{SrOH}{HTML}{c0a6d9}
\definecolor{I2}{HTML}{18186f}

\def\GeV{\,{\rm GeV}}
\def\MeV{\,{\rm MeV}}

\def\bea{\begin{eqnarray}}
\def\eea{\end{eqnarray}}

\begin{document}

\preprint{DESY-23-110}

\title{Probing an ultralight QCD axion with electromagnetic quadratic interaction}

\author{Hyungjin Kim}
\affiliation{Deutsches Elektronen-Synchrotron DESY, Notkestr. 85, 22607 Hamburg, Germany}

\author{Alessandro Lenoci}
\affiliation{Deutsches Elektronen-Synchrotron DESY, Notkestr. 85, 22607 Hamburg, Germany}

\author{Gilad Perez}
\affiliation{Department of Particle Physics and Astrophysics,\\
Weizmann Institute of Science, Rehovot, Israel 7610001}

\author{Wolfram Ratzinger}
\affiliation{Department of Particle Physics and Astrophysics,\\
Weizmann Institute of Science, Rehovot, Israel 7610001}

\begin{abstract}
The axion-gluon coupling is the defining feature of the QCD axion. 
This feature induces additional and qualitatively different interactions of the axion with standard model particles -- quadratic couplings.
Previously, hadronic quadratic couplings have been studied and experimental implications have been explored especially in the context of atomic spectroscopy and interferometry. 
We investigate additional quadratic couplings to the electromagnetic field and electron mass.
These electromagnetic quadratic couplings are generated at the loop level from threshold corrections and are expected to be present in the absence of fine-tuning.
While they are generally loop-suppressed compared to the hadronic ones, they open up new ways to search for the QCD axion, for instance via optical atomic clocks. 
Moreover, due to the velocity spread of the dark matter field, the quadratic nature of the coupling leads to low-frequency fluctuations in any detector setup. 
These distinctive low-frequency fluctuations offer a way to search for heavier axions. 
We provide an analytic expression for the power spectral density of this low-frequency background and briefly discuss experimental strategies for a low-frequency background search. 
\end{abstract}

\maketitle

\section{Introduction}\label{sec:level1}
The axion solution of the strong CP problem requires the axion field to couple to the strong sector~\cite{Peccei:1977hh, Peccei:1977ur, Weinberg:1977ma, Wilczek:1977pj, Kim:1979if, Shifman:1979if, Zhitnitsky:1980tq, Dine:1981rt}. 
Couplings between ultralight spin-0 fields and the strong sector occur also in models addressing other theoretical questions, such as the quark-flavor puzzle and electroweak hierarchy problem, or in phenomenological models like the Higgs-portal~\cite{Froggatt:1978nt, Piazza:2010ye,  Graham:2015cka, Arvanitaki:2016xds, Banerjee:2018xmn, Banerjee:2020kww, Arkani-Hamed:2020yna, TitoDAgnolo:2021nhd, TitoDAgnolo:2021pjo, Chatrchyan:2022dpy}. 
If the spin-0 field is a scalar, these couplings are severely constrained by equivalence-principle and fifth-force bounds~\cite{Adelberger:2003zx, Berge:2017ovy}. 
If the spin-0 field is a pseudo-scalar field, long-range forces do not appear at leading order, and therefore, the corresponding bounds are dramatically weaker. 
A similar trend in the strength of bounds is obtained in cases where the spin-0 field is assumed to be ultralight dark matter (ULDM); the bound on scalar DM is more than 10 orders of magnitude stronger than that of a pseudo-scalar for the same field content~\cite{Banerjee:2022sqg}. 

Axion searches are mostly based on its anomalous coupling to the photon or axial-vector couplings to standard model particles. 
Below the confinement scale, however, strong confining dynamics generate sizable quadratic couplings of the QCD axion to SM scalar operators in the strong sector\footnote{In generic axion models these are suppressed by the axion mass~\cite{Banerjee:2022sqg}}, offering new directions for axion searches.
For instance, the quadratic couplings can change the potential structure of the axion in a finite density environment, which can be examined in extreme stellar environments such as white dwarfs and neutron stars~\cite{Hook:2017psm, Balkin:2020dsr, Zhang:2021mks, Balkin:2022qer}.

Moreover, these hadronic quadratic couplings induce small time-oscillations of nuclear parameters, if the axion constitutes the observed dark matter. 
Such small oscillations of nuclear parameters can be probed by atomic spectroscopy and/or interferometry, for instance by atomic clocks~\cite{Kim:2022ype}. 
While atomic spectroscopy provides an interesting way to probe axion dark matter, it is still challenging to probe the axion via hadronic quadratic couplings since atomic clocks are in general less sensitive to the variation of nuclear parameters.

In this work, we explore another kind of quadratic interaction of the QCD axion -- the quadratic interactions with the electromagnetic field and the electron. 
The former arises from one-loop corrections, while the latter is induced by two-loop corrections. 
Although such couplings are generally smaller than their hadronic counterpart, they allow us to probe the QCD axion with a wider range of experimental setups, some of which are more sensitive. These couplings are induced and dominated by loops involving IR states, and therefore  are expected to be naturally present in the theory regardless of the details of the microscopic theory. 
The main objective of this work is, therefore, to study the interplay between current and near-future experimental sensitivity of the quadratic axion couplings to the strong and electromagnetic sectors.

\begin{figure*}[!t]
\centering
\includegraphics[width=0.7\textwidth]{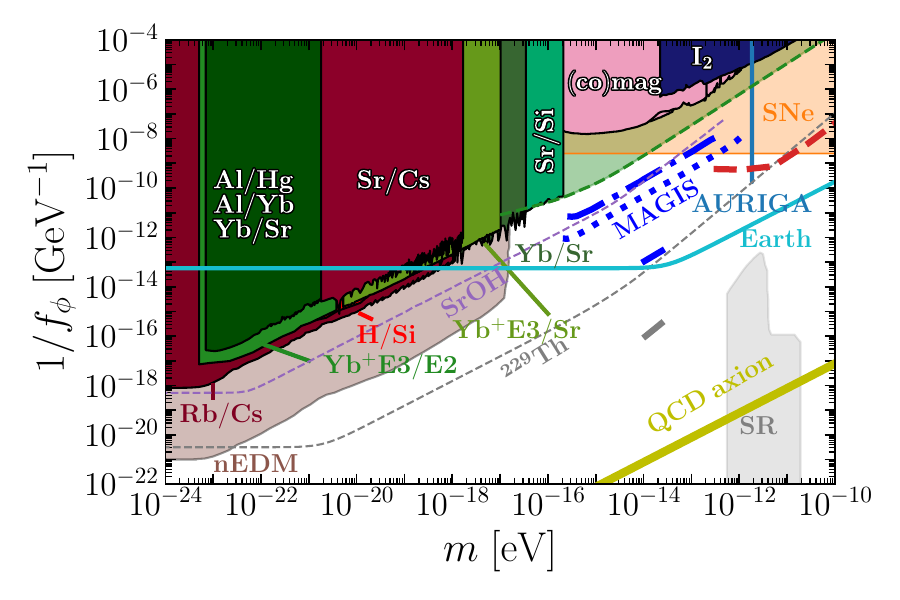}
\caption{%
Summary of constraints. 
Constraints with microwave clocks are shown in a red tone: Rb/Cs fountain clocks ({\color{RbCs} \bf Rb/Cs})~\cite{Hees:2016gop}, a H-maser with a Si cavity ({\color{HSi} \bf H/Si})~\cite{Kennedy:2020bac}, and strontium and cesium clocks ({\color{SrCs} \bf Sr/Cs})~\cite{Sherrill:2023zah}. 
They receive the dominant contribution from the variation of nuclear parameters~\cite{Kim:2022ype}. 
Searches based on optical clock transitions are shown with a green color scheme: Yb$^+$ and Sr ({\color{YbSr} \bf Yb/Sr})~\cite{Sherrill:2023zah}, Sr with a Si cavity ({\color{SrSi} \bf Sr/Si}) \cite{Kennedy:2020bac}, Al$^+$, Hg$^+$, Yb, and Sr ({\color{AlHg} \bf Al/Hg, Al/Yb, Yb/Sr})~\cite{2021Natur.591..564B}, and the electric-octupole (E3) and the electric-quadrupole (E2) transitions of Yb$^+$ ion ({\color{YbE3E2} \bf Yb$^+$E3/E2}), and Yb$^{+}$ (E3) and Sr ({\color{YbE3Sr} \bf Yb$^+$E3/Sr})~\cite{Filzinger:2023zrs}. 
The region bounded by the {\color{YbE3E2} \bf green dashed} line is excluded by comparing measured frequency uncertainties in Yb$^+$E3/E2 with the low-frequency fluctuations of the axion DM (see Section~\ref{sec:cross} for details). 
Other constraints and projections are shown as follows: (co)magnetometers ({\color{comag} \bf  pink}) with a projection of NASDUCK ({\color{comag} \bf pink dashed}) \cite{Bloch:2019lcy, Bloch:2021vnn, Lee:2022vvb, Wei:2023rzs}, molecular iodine I$_2$ spectroscopy ({\color{I2} \bf dark blue}) \cite{Oswald:2021vtc}, MAGIS-100/MAGIS-km ({\color{blue} \bf dot-dashed/dotted blue}), a projection of CASPEr-electric ({\color{red} \bf red dashed}) \cite{Aybas:2021nvn}, the AURIGA resonant bar gravitational wave experiment ({\color{auriga} \bf emerald})~\cite{PhysRevLett.118.021302}, oscillating neutron EDM ({\color{edm} \bf brown})~\cite{Abel:2017rtm}, supernova 1987A ({\color{sne} \bf orange}) \cite{Raffelt:2006cw}, axion superradiance constraints ({\color{gray} \bf gray}) \cite{Arvanitaki:2014wva}, $^{229}$Th nuclear isomer transition ({\color{gray} \bf gray dashed}), and
strontium monohydroxide SrOH ({\color{SrOH} \bf violet dashed}). 
The diagonal line in the bottom right is the minimal QCD axion line ({\color{qcd} \bf olive}), $ m^2 f_\phi^2 \simeq m^2_\pi f_\pi^2 $.
Spectroscopy bounds above the cyan solid line must be taken carefully as the axion could develop a static profile around the Earth ({\color{earth} \bf cyan})~\cite{Hook:2017psm}. 
In addition, we show the reaches of MAGIS-km and $^{229}$Th nuclear clock as a thick blue and gray line in scenarios where the DM density in the solar system is enhanced via capture processes~\cite{Budker:2023sex}.
Constraints from magnetometers and supernova are obtained by assuming axion-nucleon couplings from KSVZ-like QCD axion models~\cite{GrillidiCortona:2015jxo}.}
\label{fig:summary}
\end{figure*}

This work is organized as follows.
In Section~\ref{sec:qd_coupling}, we briefly discuss hadronic quadratic couplings from the axion-gluon interaction. 
We then show that the hadronic quadratic couplings generate quadratic interactions with the photon and the electron at one and two-loop levels, respectively. 
In Section~\ref{sec:spec}, we discuss the implications of electromagnetic quadratic interactions in axion searches with atomic spectroscopy and gravitational wave detectors. 
In Section~\ref{sec:cross}, we discuss in more detail the signal spectrum of axion dark matter generated by quadratic interactions. We show that the quadratic nature of the coupling leads to low-frequency stochastic fluctuations of observables besides the coherent harmonic signals at frequencies corresponding to two times the axion mass. 
We further discuss possibilities to constrain and probe such stochastic signals in an experimental setup with a single detector and multiple detectors.
We conclude in Section~\ref{sec:conclusion}.
We use natural units $c=\hbar=1$ throughout this work.

\section{Quadratic couplings}\label{sec:qd_coupling}
We start from the axion coupling to the standard model gluon field, 
\bea
{\cal L} = \frac{g_s^2}{32\pi^2} \frac{\phi}{f_\phi} G^a_{\mu\nu} \widetilde G^{a\mu\nu},
\label{axion_gluon}
\eea
where $f_\phi$ is the axion decay constant, $g_s$ is the strong coupling, $G^a_{\mu\nu}$ and $\widetilde{G}^{a}_{\mu\nu}$ are the gluon field strength and its dual. 
We do not take into account any other couplings in this work; i.e. we consider KSVZ-like models where axion couplings to the axial vector currents of SM fields are absent at UV scales. 
Model-dependent couplings will not change our analysis, but they may lead to additional bounds on the axion parameter space. 

The axion-gluon coupling \eqref{axion_gluon} naturally leads to hadronic quadratic couplings below the QCD scale. 
For instance, the pion mass can be found from the chiral Lagrangian as $m_\pi^2(\theta) = B (m_u^2 + m_d^2 + 2 m_u m_d \cos \theta)^{1/2}$ with $B = -\langle \bar q q\rangle_0/f_\pi^2$ and $\theta = \phi/f_\phi$~\cite{DiVecchia:1980yfw}. 
Expanding the pion mass around $\theta=0$, we find a quadratic coupling to pions, ${\cal L}_{} \supset \theta^2 \pi^2$. 
Furthermore, the nucleon mass depends on the pion mass through ${\cal L} \supset 4c_1 m_\pi^2(\theta) \bar N N$ with $c_1 = -1.1\GeV^{-1}$~\cite{Hoferichter:2015tha}, which leads to a quadratic interaction between the axion and nucleons as well. 

The hadronic quadratic interactions introduce time oscillations of the nuclear parameters if the axion is the dark matter in the present universe.
As it is clear from the chiral Lagrangian, the pion and nucleon mass will receive a small oscillating component, e.g.
\bea
\frac{\delta m_\pi^2}{m_\pi^2}
= - \frac{z}{2(1+z)^2} \theta^2(t)
\eea
with $z = m_u/m_d$. The amplitude of the oscillating $\theta^2(t)$ is proportional to the dark matter density $\rho\approx \theta^2(t)m^2 f_\phi^2$, where $m$ denotes the axion mass.
Other nuclear parameters that depend on the pion mass, such as the nuclear $g$-factor will also receive such a time-oscillating component.
As a consequence, atomic energy levels oscillate and, in addition, any object whose mass receives QCD contributions experiences an acceleration due to the axion dark matter background. 

Based on this observation, it was shown in Ref.~\cite{Kim:2022ype} that spectroscopic and interferometric measurements, such as atomic clocks and gravitational wave interferometers, can be used to search the axion at the low mass range. 
In particular, clock comparison tests using hyperfine transitions are considered since hyperfine transitions are directly affected by the variation of nuclear parameters.
In addition, gravitational wave interferometers are considered as most mass of the test bodies comes from QCD, and therefore, they fluctuate inevitably in the axion dark matter background. 

As discussed in the introduction, the hadronic interactions also lead to EM quadratic couplings at low energy scales through loop corrections. 
Below, we detail how these EM couplings are induced and show that all of these effects are due to the variation of the pion mass. 

\subsection{Quadratic interaction with the electromagnetic field}
We first consider the quadratic coupling to the electromagnetic field,
\bea
{\cal L} = 
-\frac{1}{4} F_{\mu\nu} F^{\mu\nu}
+ \frac{C_{\gamma}}{4} \frac{\phi^2}{f_\phi^2} F_{\mu\nu} F^{\mu\nu}
\label{C_gamma_Lagrangian}
\eea
at energy scales below the pion mass. 
The coefficient $C_{\gamma}$ is given as
\bea
C_\gamma =  - \frac{z}{24(1+z)^2} 
\frac{\alpha}{\pi} 
\bigg(1 + 8 \frac{\sigma_{\pi N}}{m_N} \bigg) 
\simeq - 3\times 10^{-5}. 
\label{C_gamma}
\eea
Here $z = m_u / m_d \simeq 0.46$ and $\sigma_{\pi N} = \partial m_N / \partial \ln m_\pi^2 \sim {\cal O}(50)\MeV$. 
The coefficient $C_\gamma$ can be directly obtained from the one-loop computation of $\phi \phi \to \gamma\gamma$ via a pion loop or a nucleon loop. 
Alternatively, it can be read off from the threshold correction to the running of the fine structure constant with respect to the variation of the pion and the nucleon masses. 

Consider the running of the electromagnetic coupling $1/e^2$ from $\Lambda_{\rm UV}$ to $\Lambda_{\rm IR}$. 
Let us assume a single charged particle whose mass $m_\pi$ is in between these scales. 
The gauge coupling runs as $1/e^2(\Lambda_{\rm IR}) - 1/e^2(\Lambda_{\rm UV}) = \int_{\Lambda_{\rm IR}}^{\Lambda_{\rm UV}} d\ln\mu \, [2\beta_e(\mu)/e^3]$ where $\beta_e(\mu) = (b e^3/8\pi^2)$, $b = \frac{2}{3} \sum_f Q_f^2 + \frac{1}{6} \sum_s Q_s^2$, and the sums over $f$ and $s$ account for Dirac fermions and complex scalars, respectively. 
For a fixed UV value of the gauge coupling, one finds that the gauge coupling at low energy depends on the mass of the charged particle,
\bea
\delta
\bigg(  \frac{1}{e^2_{\rm IR}} \bigg)
=  - \frac{\Delta b}{4\pi^2}\frac{\delta m_\pi}{m_\pi}.
\eea
Here $\Delta b$ is the change of the beta function coefficient at the threshold; $\Delta b= 2Q^2/3,\ Q^2/6$ for a Dirac fermion and a complex scalar field of charge $Q$, respectively.
In the effective Lagrangian, this dependence is incorporated by  
$$
{\cal L} = \frac{\Delta b e^2}{(4\pi)^2} \frac{\delta m_\pi}{m_\pi} F_{\mu\nu}F^{\mu\nu}.
$$
From this, we obtain
\begin{align}
\frac{\delta\alpha}{\alpha}
= C_\gamma \theta^2
= \frac{\alpha}{\pi} \sum_i \Delta b_i \frac{\delta m_i}{m_i}
= \frac{\alpha}{12\pi}
\left[ 1 + \frac{8 \sigma_{\pi N}}{m_N} \right] \frac{\delta m_\pi^2}{m_\pi^2},
\label{alpha_fluc}
\end{align}
where we include the variation of the nucleon mass; the fine structure constant fluctuates as the pion mass changes due to the background axion dark matter.

\subsection{Quadratic interaction with electron mass}
Furthermore, the hadronic quadratic couplings lead to a quadratic coupling to the electron mass,
\bea
{\cal L} = - C_e m_e \frac{\phi^2}{f_\phi^2} \bar e e ,
\eea
where the coefficient $C_e$ is 
\bea
C_e \simeq \frac{3 \alpha}{4\pi} C_\gamma  \ln \frac{m_\pi^2}{m_e^2}. 
\label{C_e}
\eea
This effect arises at two-loop order; it is suppressed by $(\alpha/\pi)^2$. 

We estimate the coefficient $C_e$ in the following way. 
Below the QCD scale, the dim-$6$ operator \eqref{C_gamma_Lagrangian} contributes to the running of the electron mass.
In QED, the correction to the electron mass due to running from the QCD scale to the electron mass is given by $(\delta m_e/m_e) = (3 \alpha / 4\pi) \ln m_\pi^2 / m_e^2$. 
Since $\theta^2 (t)$ oscillates at a frequency much smaller than the electron mass or the QCD scale, we can effectively take $C_\gamma\theta^2$ as a constant and absorb it by rescaling the gauge field $A_\mu \to (1+ C_\gamma\theta^2)^{1/2} A_\mu$. 
This is equivalent to taking $e^2 \to e^2 (1 + C_\gamma \theta^2)$. 
Using the QED result, one finds that the dim-$6$ operator contributes to the running as
$$
\frac{\delta m_e}{m_e}
\simeq \frac{3 \alpha}{4\pi} C_\gamma \theta^2 \ln (m_\pi^2/m_e^2)
= C_e \theta^2
$$
from which we estimate $C_e$ as in \eqref{C_e}. 
An explicit diagrammatic computation leads to the same result. 
We do not consider the variation of electron mass further, however, as its effect on observables is usually much smaller than the variation of the fine structure constant and nuclear parameters. 

\section{Implications}\label{sec:spec}
The quadratic couplings to the electromagnetic field and the electron mass offer alternative ways to search for the QCD axion.
Previously, Ref.~\cite{Kim:2022ype} focused on the quadratic coupling to hadrons.\footnote{
A generic ultralight dark matter model with $\phi^2$ coupling was considered in~\cite{Stadnik:2015kia, Stadnik:2016zkf}.}
Assuming that the axion constitutes DM, it was pointed out that those lead to variations of the nucleon mass and nuclear $g$-factor over time and that atomic clocks based on hyperfine transitions could pick up the axion DM-induced signals.
When additionally taking into account quadratic couplings to the electromagnetic field and the electron mass, a wider range of experiments becomes sensitive, including atomic clocks based on electronic transitions. 
Since such optical clocks usually have a shorter averaging time and better sensitivities, one can expect to possibly probe a wider range of parameter space.

For clarification, let us briefly recall the argument form Ref.~\cite{Arvanitaki:2014faa}, how a clock comparison test is sensitive to a coupling of ultralight dark matter like e.g. the one introduced in \eqref{alpha_fluc}. 
Consider two stable frequency standards $f_A$ and $f_B$. 
Suppose that each frequency standard has a slightly different dependence on the fine structure constant, $f_A \propto \alpha^{\xi_A}$ and $f_B \propto \alpha^{\xi_B}$ with $\xi_A \neq \xi_B$. 
Due to the time-variation of the fine structure constant caused by the DM field, the ratio of these two frequencies fluctuates as
$$
\frac{\delta(f_A/f_B)}{(f_A/f_B)} = (\xi_A -\xi_B) \frac{\delta \alpha}{\alpha} 
\propto \theta^2(t) ,
$$
where $\theta^2(t) = (\rho_0 / m^2 f_\phi^2 )\cos(2mt)$ is related to $\delta \alpha/\alpha$ by \eqref{alpha_fluc}. 
By monitoring the frequency ratio and investigating if the time series contains any harmonic signal at $\omega =2m$, one can probe the QCD axion.

Generically the fractional frequency deviation arising from a quadratic coupling can be written as
\bea
\frac{\delta f_A}{f_A} = K_A \theta^2 (t),
\eea
where $K_A$ is the sensitivity coefficient that depends on the atomic species and transition.
It takes all the effects (hadronic and electromagnetic) into account. 
A list of the coefficient $K_A$ for different atom species is available in Appendix~\ref{app:sen}.

Any stable frequency standard can be used to search for the QCD axion.
Ref.~\cite{Kim:2022ype} only used hyperfine transitions as only hadronic quadratic couplings were considered in that work and hadronic couplings do not affect the electronic transition to leading order. 
Possible variations of electronic transition caused by oscillations of the nuclear charge radius were investigated in Ref.~\cite{Banerjee:2023bjc}. 
Due to the electromagnetic quadratic couplings described above, electronic transition levels now change directly as the background DM oscillates. 
Although these new quadratic couplings are at least one-loop suppressed, they still lead to competitive bounds compared to microwave clocks as optical clocks have orders of magnitude smaller frequency uncertainties.
Also, due to a smaller required averaging time of the spectroscopic measurement, optical clocks probe a slightly higher mass range than microwave clocks as shown in Figure~\ref{fig:summary}. 

The QCD axion also changes the length of objects through its electromagnetic quadratic couplings. 
Since the length of any object is proportional to the size of its atoms, i.e. the Bohr radius, $L \propto (m_e\alpha)^{-1}$, the QCD axion induces a strain $\Delta L / L = - (\delta \alpha/ \alpha + \delta m_e/m_e)$. 
This small strain can be effectively probed by resonant bar gravitational detectors, such as AURIGA~\cite{PhysRevLett.118.021302}, which is also shown in Figure~\ref{fig:summary}. 
We also present the projected sensitivities of the atom interferometry MAGIS-100 and MAGIS-km experiments \cite{Aybas:2021nvn}. Atom interferometers compare the phase accumulated by two delocalized atom clouds. The presence of a background ULDM field can affect the phase of the atomic cloud in two ways: (i) via the value of the internal energy splitting, like an atomic clock~\cite{Arvanitaki:2016fyj}; (ii) via the exertion of an additional force on the atomic clouds, causing acceleration~\cite{Graham:2015ifn}.

\section{Spectrum}\label{sec:cross}
The quadratic axion DM signal discussed so far is the harmonic signal, $s(t) \propto \theta^2(t) \propto \cos(2mt)$, at the frequency twice the dark matter mass, $\omega =2m$. 
By investigating if the detector output has an oscillating component at $\omega =2m$ via matched filtering, it is possible to probe or constrain interactions of ULDM with SM particles.

The quadratic operator exhibits not only coherent harmonic oscillations but also distinctive low-frequency stochastic fluctuations at $\omega \lesssim m v^2$, where $v$ denotes the DM velocity. 
This offers another opportunity to test the QCD axion. 
Recently, Masia-Roig et al~\cite{Masia-Roig:2022net} showed that a network of sensors can be used to probe such low-frequency stochastic background in the context of non-gravitational quadratic interactions of ULDM with SM particles. 
Flambaum and Samsonov~\cite{Flambaum:2023bnw} argued that, by directly comparing the low-frequency background with experimentally measured uncertainties, it is possible to set limits on the QCD axion parameter space at higher masses.

We provide below the analytic spectrum of the low-frequency fluctuation of the axion dark matter from its quadratic interactions and project the sensitivity of different detector networks. 

To see how this low-frequency stochastic noise arises from the quadratic operator, let us assume that the signal is proportional to the quadratic operator as follows,
$$
s(t) = K \theta^2 (t),
$$
with arbitrary constant $K$.
Once we expand the field as\footnote{See Appendix~\ref{app:qd_spec} and Ref.~\cite{Kim:2021yyo, Kim:2023pkx} for more detailed discussions on this statistical description of wave dark matter.}
\bea
\phi(t,x) = \sum_i \frac{1}{\sqrt{2mV}}
\left[ \alpha_i e^{-i k_i \cdot x} 
+ \alpha_i^* e^{ i k_i \cdot x }
\right]
\eea
with complex random numbers $(\alpha_i, \alpha_i^*)$, it is clear that the quadratic operator contains the sum, $\omega_i + \omega_j$, and the difference, $\omega_i - \omega_j$ of two frequencies in the field,
$$
\phi^2(t,0) \supset
\alpha_i \alpha_j e^{- i(\omega_i + \omega_j) t}
+\alpha_i \alpha_j^* e^{- i(\omega_i - \omega_j) t}
+ {\rm h.c.}.
$$
In the non-relativistic limit, the first term $\omega_i + \omega_j \simeq 2m$ provides the harmonic signal at $\omega =2m$.
The second term, on the other hand, provides a low-frequency fluctuation at $\omega \lesssim mv^2$.

A more careful investigation is possible via the power spectrum of the quadratic operator. 
The one-sided power spectral density (PSD) of the signal, $P_s(f)$, is defined as
\bea
\langle \tilde s(f) \tilde s^*(f')
\rangle
= \delta(f - f') \frac{1}{2} P_s(f),
\eea
where $s(t) = \int df\, e^{-2\pi i f t}\, \tilde s(f)$. 
Following Ref.~\cite{Kim:2023pkx}, for a normal DM velocity distribution $n(\vec v) = [(\rho_0 / m) / (2\pi \sigma^2)^{3/2}] \exp(-v^2/2\sigma^2)$ with the mean dark matter density $\rho_0$ and the velocity dispersion $\sigma$, one finds the signal PSD as
\bea
P_s(f) 
= K^2 \frac{ \theta_0^4 }{4}  \tau_\phi
\Big[ A(f) + B(f) \Big]
\label{signal_spec}
\eea
where $\tau_\phi = 1 / m \sigma^2$ is the coherence time and
\bea
A(f ) &=&
\pi \frac{\bar v^4}{\sigma^4} e^{-\bar v^2/\sigma^2} \theta(\bar v^2)
\\
B(f) &=&
4\bar\omega K_1( \bar\omega). 
\label{B_ISO}
\eea
Here $\bar v^2 = 2\pi f /m -2$, $\bar\omega =2\pi f / m\sigma^2$, $K_n(x)$ is the modified Bessel function of the second kind, and $\theta(x)$ is the step function.
The expression shows two distinctive frequency components: $A(f)$ represents the harmonic signal at $\omega=2\pi f=2m$, and $B(f)$ represents the low-frequency stochastic fluctuation. 
For a detailed derivation, see Appendix~\ref{app:qd_spec}. 

The low-frequency stochastic background behaves similarly to white noise and is therefore difficult to distinguish from other random noises in a detector.\footnote{It might be possible to disentangle axion DM-induced fluctuation from Gaussian random noise since axion DM fluctuations from quadratic interaction follows the exponential distribution rather than the normal distribution~~\cite{Flambaum:2023bnw}.}
If it is somehow possible to arrange the output data in a way that it is insensitive to the axion DM signal, then it could be possible to calibrate the noise and therefore detect the axion with only one experiment. 
Alternatives are the following two approaches: (i) the reported stability of clocks can be used to constrain the parameter space as the low-frequency stochastic DM background would lead to larger fluctuations than the ones observed; (ii)
to possibly detect the axion, one relies on the cross-correlation between multiple experiments for which the individual experiment-intrinsic noise cancels while the axion signal persists. 
In the following, we discuss these two aspects in more detail.

\subsection{Single detector setup}
As already demonstrated in Ref.~\cite{Flambaum:2023bnw}, by comparing the low-frequency fluctuations with the measured uncertainty of clocks, one can place lower limits on the decay constant $f_\phi$. In a repeated measurement of a given frequency standard, there will be varying fluctuations due to the experiment's intrinsic effects as well as possibly the axion signal. Since the low-frequency part of the signal has very similar properties to white noise, we expect the two to be hardly distinguishable.
Even if we consider the axion signal as just another component of the noise, we can still constrain the axion by requiring that the noise due to the axion is smaller than the total observed one.

To illustrate this further, we choose the measurement of the frequency ratio of Yb$^{+}$ electric-octupole (E3) and electric-quadrupole (E2) clock transitions~\cite{Filzinger:2023zrs}.
As explained above, one way to extract the constraints on $1/f_\phi$ is to directly compare the low-frequency noise $P_s \sim K^2 \theta_0^2 \tau_\phi$~\eqref{signal_spec} with the measured clock frequency uncertainties.
The fluctuations of the measured frequency ratio in Ref.~\cite{Filzinger:2023zrs} are consistent with white noise of $P_n(f) = \sigma_n^2 \simeq (10^{-14}/\sqrt{\rm Hz})^2$. 
A direct comparison leads to the constraint on $1/f_\phi$ as $f_\phi^{-1} = [m^2 \sigma_n/ (2 K \rho_0 \sqrt{\tau_\phi} )] ^{1/2}$ with $K \simeq 10^{-4}$. 

More precisely, we compute the Allan deviation caused by the axion DM and compare it to the experimentally reported value.
The Allan deviation is defined in~\eqref{allan_def} and the expected value for the axion DM is provided in Appendix~\ref{app:allan}. 
We find
\bea
\frac{1}{f_\phi}
= 
\left[
\frac{m^4 \sigma_{n}^2(\tau)}{8 K^2 \rho_0^2 {\cal I}(\tau/ 2 \tau_\phi)}
\right]^{1/4}
\eea
where $\sigma_{n}(\tau)$ is the reported Allan deviation with an averaging time $\tau$. 
The detailed derivation and the function ${\cal I}(x)$ are given in Appendix~\ref{app:allan}. 
This constraint is shown by the green dashed region in Figure~\ref{fig:summary} and \ref{fig:cross}.

\subsection{Multi-detector setup}
If two or more detectors are available, it is possible to distinguish the axion DM signal from the detector's noise by cross-correlating multiple detector outputs. 
Suppose we have two detector outputs $d_{1,2} (t) = s_{1,2}(t) + n_{1,2}(t)$.
If we now consider the correlation between the two outputs $\langle d_1 d_2\rangle$, we expect the noises in the two detectors to be uncorrelated amongst themselves, $\langle n_1 n_2 \rangle \sim 0$, while the signal is $\langle s_1 s_2\rangle \neq 0 $ as long as the two detectors are placed within one coherence length $L < \lambda \approx 1/(mv)$.
In practice, this is done by constructing an observable as $Y = \int dt \int dt' \, s_1(t) s_2(t') Q(t-t')$ with some real filter function $Q(t-t')$.
The signal and noise are computed as $S = \langle Y \rangle$ and $N^2 = [\langle Y^2 \rangle - \langle Y \rangle^2]_{s=0}$, respectively.
The maximum signal-to-noise ratio is~\cite{Maggiore:2007ulw}
\bea
\frac{S}{N} = 
\left[
2 T \int_{f_l}^{f_u} df \, \frac{|P_{\rm cross}|^2}{P^2_n(f)}
\right
]^{1/2}
\eea
where $f_{u,l}$ is the highest and lowest frequency where $P_n(f)$ is available, $T$ is the total observation time scale, $P_n(f) = [ P_{n_1}(f) P_{n_2} (f) ]^{1/2}$ is the noise PSD, and $P_{\rm cross}$ is the cross-correlation defined as 
\bea
\langle \tilde s_1 (f) \tilde s_2^*(f') \rangle = \delta(f-f') \frac{1}{2} P_{\rm cross}(f). 
\eea
For $N_{\rm det}$ detectors, the above expression is modified as $T \to [N_{\rm det}(N_{\rm det} - 1)/2] T$ assuming that the noise PSD in all detectors is more or less the same.

The cross-correlation PSD from the axion DM can be computed straightforwardly with the formulation described above. 
For a normal velocity distribution with zero mean velocity, we find
\bea
P_{\rm cross}(f, \vec L)
= K_1 K_2  \frac{\theta_0^4\tau}{4}
B_{\rm cross}(f, \vec L)
\eea
where
\begin{align}
B_{\rm cross}(f, \vec L) =
2\int_{-\infty}^\infty dx
\frac{e^{-i \bar \omega x}}{(1+x^2)^{3/2}}
\exp\left[
- \frac{(m\sigma L)^2}{1+x^2}
\right] . 
\end{align}
Note that $K_{1,2}$ is the sensitivity coefficient defined as $s_i = K_i \theta^2(t)$ and $L$ is the detector separation. 
The detailed derivation and more general expressions with dark matter mean velocity are given in Appendix~\ref{app:qd_spec}. 
Note that the above expression coincides with \eqref{B_ISO} in the $L\to 0$ limit. 

\begin{figure}
\centering
\includegraphics[width=0.45\textwidth]{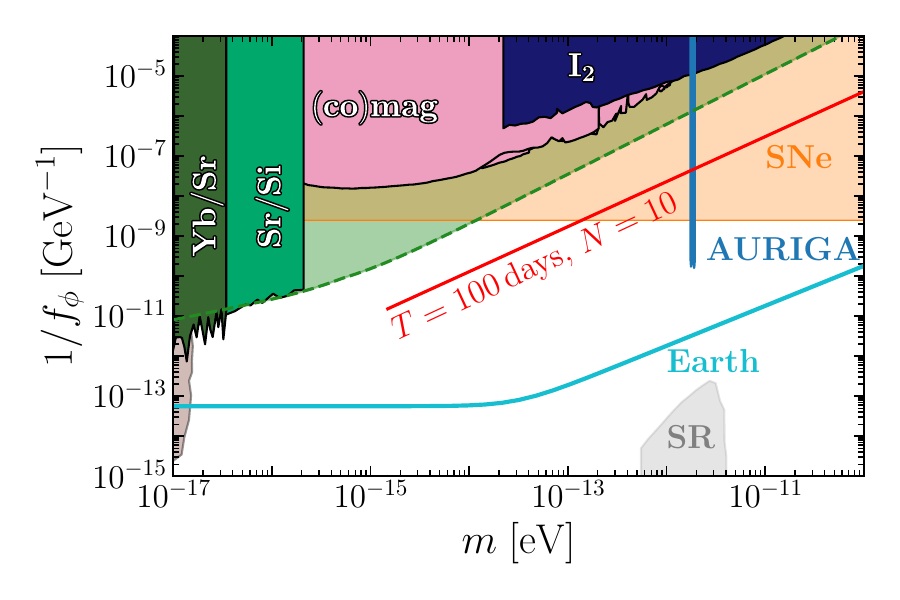}
\caption{A projection for cross-correlation with optical clock systems (red line). 
We choose $T=100\,{\rm days}$, $N_{\rm det}=10$ detectors, and $S/N$=3.}
\label{fig:cross}
\end{figure}

In Figure~\ref{fig:cross}, we choose optical clock systems to investigate to which extent they can probe the QCD axion parameter space at a higher mass range. 
Assuming only white noise and $m\sigma L \ll1$ such that $B_{\rm cross} \approx B$, one finds the projected sensitivity on $1/f_\phi$ as 
\bea
\frac{1}{f_\phi}
\approx
\left[ 
\frac{P_n}{4 K_1 K_2} \frac{S}{N} \frac{m^4}{\rho_0^2}
\right]^{\frac14}
\left[
\frac{\pi}{T \tau_\phi \min(1, 2 \pi f_u \tau_\phi)}
\right]^{\frac18}
\eea
We choose $K_{1,2} = 10^{-4}$, measurement frequency $f_u=1\,{\rm Hz}$, and $P_n^{1/2}(f) = 10^{-16}/\sqrt{\rm Hz}$.
Unlike the single detector setup in the previous section, the signal-to-noise ratio and the projection on $1/f_\phi$ show a mild improvement as a function of observation time and the number of detectors. 
This can be seen by comparing the projection of a network with $T=100\,{\rm days}$ and $N_{\rm det}=10$ detectors shown in Fig.~\ref{fig:cross} as a red line, with the green dashed region showing the Yb$^+$ (E3)/(E2) constraint from the previous section in a single detector setup. Crucially the multi-detector setup allows for the detection of the axion since the non-vanishing cross-correlation can distinguish the signal from detector noise.

\section{Conclusion}\label{sec:conclusion}
In this work, we have considered the quadratic interactions of the QCD axion with the electromagnetic field and the electron mass.
These quadratic interactions naturally arise as long as the axion couples to the gluon field of the standard model. 
Similar to the quadratic interaction with pions and nucleons, such interactions lead to oscillating atomic energy levels.
Contrary to the hadronic coupling, the electromagnetic interaction directly affects the electronic energy levels, making systems that depend on these energy levels sensitive to axion DM. 
As examples of such systems, we studied optical clocks, resonant-bar gravitational wave detectors, and atom interferometers.

We have summarized existing constraints and projected sensitivities of future nuclear clocks in Fig. \ref{fig:summary}. 
While they are still far from the minimal QCD axion parameter space, they provide alternative ways to search the QCD axion. 
Moreover, the quadratic nature inevitably introduces a low-frequency stochastic background. 
We have derived an analytic expression for the low-frequency spectrum of the ultralight DM-induced signal.
By directly comparing the axion DM-induced low-frequency fluctuations with measured clock uncertainties, we show that the Yb$^+$ (E3) and (E2) comparison can also probe heavier axions than those considered in previous work~\cite{Filzinger:2023zrs}.
In addition, with several assumptions, we have also projected the sensitivity of a network of detectors, which could probe this higher mass range further.

\begin{acknowledgments}
We would like to thank Abhishek Banerjee for useful discussions and for providing us with tabulated data of the Yb$^{+}$ comparison test. 
The work of HK and AL was supported by
the Deutsche Forschungsgemeinschaft under Germany’s
Excellence Strategy - EXC 2121 Quantum Universe
- 390833306.
The work of GP is supported by grants from BSF-NSF, Friedrich Wilhelm Bessel research award of the Alexander von Humboldt Foundation, ISF, Minerva, SABRA - Yeda-Sela - WRC Program, the Estate of Emile
Mimran, and the Maurice and Vivienne Wohl Endowment.
\end{acknowledgments}

\bigskip
{\noindent\it Note added:} While this work was being finalized, a related work~\cite{Beadle:2023flm} appeared on the arXiv, which shares some of the points discussed above.

\appendix

\section{Clock comparison test}\label{app:sen}
We list the sensitivity coefficient for the QCD axion for different atomic species. 
Let us consider the frequency standards based on hyperfine and electronic transitions.
The transition frequency is parameterized as
\begin{align}
f_{\rm hfs} &= g \frac{m_e^2}{m_p} \alpha^4 F_{\rm hfs}(\alpha) ,
\\
f_{\rm elec} &= m_e \alpha^2 F_{\rm elec} (\alpha),
\end{align}
where $g$ is the nuclear $g$-factor, and $F(\alpha)$ is the relativistic correction.

There are total 4 parameters, $\{g, m_e, m_p, \alpha \}$.
Each of them varies in time. 
The transition frequency can be conveniently written as $f_A = g_A^{K_g} m_e^{K_{m_e}} m_p^{K_{m_p}} \alpha^{K_{\alpha}}$. 
Since the effect of the QCD axion always arises through the variation of pion mass, the fractional frequency change can be written as
\bea
\frac{\delta f_A}{f_A}
= 
\sum_{i }
K_i \frac{\partial \ln A_i}{\partial \ln m_\pi^2}
\frac{\delta m_\pi^2}{m_\pi^2}
\eea
where the index runs over all four parameters. 
$K_{g} = 1,0$, $K_{m_e} =2,1$, and $K_{m_p} = -1,0$ for hyperfine and electronic transition, respectively. 
The values for $K_{\alpha}$ can be found in Refs.~\cite{Flambaum:2006ip, Safronova:2017xyt}. 
The dependence of each parameter on the pion mass is
\begin{align}
\frac{\partial \ln m_p}{\partial \ln m_\pi^2}
&=
\frac{\sigma_{\pi N}}{m_N}
\simeq 0.06
\\
\frac{\partial \ln \alpha}{\partial \ln m_\pi^2}
&=
\frac{\alpha}{12\pi}
\left(
1 + \frac{8 \sigma_{\pi N}}{m_N} 
\right)
\simeq
3\times 10^{-4}
\\
\frac{\partial \ln m_e}{\partial \ln m_\pi^2}
&=
\frac{3\alpha^2}{48\pi^2}
\left(
1 + \frac{8 \sigma_{\pi N}}{m_N} 
\right)
\ln \frac{m_\pi^2}{m_e^2}
\simeq 6 \times 10^{-6}.
\end{align}
where $\sigma_{\pi N} = \partial m_N / \partial \ln m_\pi^2$. 
For the $g$-factor, one finds 
$\partial \ln g_p / \partial \ln m_\pi^2 = -(g_A^2/g_p) [m_N m_\pi/(8\pi f_\pi^2)] \simeq - 0.17$
for the hydrogen atom, $\partial \ln g_{\rm Rb} / \partial \ln m_\pi^2 = -0.024$ for $^{87}$Rb, and  $\partial \ln g_{\rm Cs} / \partial \ln m_\pi^2 = 0.011$ for $^{133}$Cs~\cite{Kim:2022ype}.
For the nuclear clock transition in $^{229}$Th, the hadronic quadratic coupling is dominant, $\delta f_A/f_A \simeq (2\times 10^5) \times \delta m_\pi^2/m_\pi^2$~\cite{Kim:2022ype}. 

The above expression can be written in a more compact form:
\bea
\frac{\delta f_A}{f_A} = K_A \theta^2 . 
\eea
The sensitivity coefficient $K_A$ for each atom is listed in Table~\ref{tab:sensitivity_coeff}. 
The sensitivity coefficient of the frequency ratio of any pair of atomic transitions is simply the difference of the two respective sensitivity coefficients.

\begin{table}[t]
\centering
\begin{tabular}{c|c|c}
System
& Transition 
& $K_A$
\\ \hline\hline
H 
& Ground state hyperfine
& $+1.2\times 10^{-2}$
\\
Cs 
& Ground state hyperfine
& $+2.6\times 10^{-3}$
\\
Rb 
& Ground state hyperfine
& $+4.5\times 10^{-3}$
\\
Si 
& cavity
& $-1.5\times 10^{-5}$
\\
Sr 
& $^1S_{0} \to\, ^3P_{0}$
& $-3.2\times 10^{-5}$
\\
Al$^+$ 
& $^1S_{0} \to\, ^3P_{0}$
& $-3.1\times 10^{-5}$
\\
Hg$^+$ 
& $^2S_{1/2} \to\, ^2D_{5/2}$
& $+1.4\times 10^{-5}$
\\
Yb 
& $^1S_{0} \to\, ^3P_{0}$
& $-3.5\times 10^{-5}$
\\
Yb$^+$ (E2) 
& $^2S_{1/2} \to \,^2D_{3/2}$
& $-4.6\times 10^{-5}$
\\
Yb$^+$ (E3) 
& $^2S_{1/2} \to\, ^2F_{7/2}$
& $+6.0\times 10^{-5}$
\\
Th
& nuclear 
& $-2.2\times 10^4$
\\
\hline
\end{tabular}
\caption{Table for the sensitivity coefficient $K_A$ for the QCD axion.
The electromagnetic quadratic interactions provide the dominant effect for the optical clock transitions, while the hadronic couplings provide the dominant effects for hyperfine and nuclear clock transitions.}
\label{tab:sensitivity_coeff}
\end{table}

\section{Quadratic spectrum}\label{app:qd_spec}
Here we provide a detailed computation of the low-frequency power spectrum, following Ref.~\cite{Kim:2023pkx}. 

We expand the field as
\bea
\phi(t,x) = \sum_i \frac{1}{\sqrt{2mV}}
\left[ \alpha_i e^{-i k_i \cdot x} 
+ \alpha_i^* e^{ i k_i \cdot x }
\right].
\eea
Here $(\alpha_i ,\alpha_i^*)$ are complex random numbers.
The underlying probability distribution of this complex random number is given by~\cite{Derevianko:2016vpm, Foster:2017hbq, Centers:2019dyn, Kim:2021yyo}
\bea
p(\alpha_i) 
= \frac{1}{\pi n_i}
\exp\left[ - \frac{|\alpha_i|^2}{n_i} \right],
\label{p_a}
\eea
where $n_i$ is the mean occupation number of the mode $i$. 
In this description, the field $\phi$ is a Gaussian random field. 

The mean occupation number $n_i$ is given by the dark matter velocity distribution.
For simplicity, we assume a normal distribution
\bea
n(\vec v) = \frac{\rho_0/m}{(2\pi \sigma^2)^{3/2}}
\exp\left[
- \frac{(\vec v - \vec v_0)^2}{2\sigma^2}
\right]
\label{normal_dist}
\eea
where $\rho_0$ is the mean dark matter density, $\vec v_0$ is the velocity of the dark matter wind relative to the experiment, and $\sigma$ is the velocity dispersion. 

We focus on the case where the signal in the detector is of the following form:
\bea
s(t) = K \theta^2(t)~,
\eea
where $K$ is a sensitivity coefficient and $\theta = \phi/f_\phi$. 
The power spectral density $P_s(f)$ is defined as
\bea
\langle 
\tilde s(f) \tilde s^{*}(f')
\rangle
= \delta(f-f') \frac{1}{2} P_s(f). 
\eea
We choose the following convention for the Fourier transformation, $s(t) = \int df e^{-2\pi i f t} \tilde s(f)$. 

The signal power spectral density is related to the PSD of the axion field as
\bea
P_s(f) = K^2 P_{\delta \theta^2} (f)
\eea
where
\bea
\langle \widetilde{\delta \theta^2} (f) 
\widetilde{\delta \theta^2}^{*} (f') 
\rangle
= \delta(f-f') \frac{1}{2} P_{\delta\theta^2}(f). \label{eq:definition_spectrum}
\eea
We have introduced $\delta\theta^2 = \theta^2 - \langle \theta^2\rangle$. 
This subtracts an unobservable constant shift in $\theta^2$. 
Note that the above power spectrum is one-sided; we only consider $f\geq 0$.

\subsection{Power spectral density}
The Fourier component of the quadratic operator is
\begin{align}
\widetilde{\theta^2}(\omega)
&=
\frac{1}{f_\phi^2} 
\frac{1}{2mV} 
\sum_{i,j} 
\nonumber\\
\times& \Big[
\alpha_i \alpha_j e^{i (\vec k_i + \vec k_j)\cdot \vec x}
(2\pi ) \delta(\omega - \omega_i - \omega_j)
\nonumber\\
& + \alpha_i \alpha_j^* e^{i (\vec k_i -\vec  k_j)\cdot \vec x}
(2\pi) \delta(\omega - \omega_i + \omega_j) 
\nonumber\\
& + \alpha_i^* \alpha_j e^{-i (\vec k_i -\vec  k_j)\cdot \vec x}
(2\pi) \delta(\omega + \omega_i - \omega_j) 
\nonumber\\
& + \alpha_i^* \alpha_j^* e^{-i (\vec k_i + \vec  k_j)\cdot \vec x}
(2\pi) \delta(\omega + \omega_i + \omega_j) 
\Big]
\end{align}
To compute $\langle \widetilde{\delta\theta^2}(\omega) \widetilde{\delta\theta^2}^*(\omega')\rangle$, the following expression is useful
\bea
\langle \alpha_i \alpha_j \alpha^*_k \alpha^*_l \rangle
= n_i n_j ( \delta_{ik} \delta_{jl} + \delta_{il} \delta_{jk} ) . 
\eea
The angle bracket denotes an ensemble average, defined as
\bea
\langle {\cal O} \rangle
= \int \bigg[ \prod_{i} d^2\alpha_i \, p(\alpha_i) \bigg ] {\cal O} . 
\eea
After a straightforward computation, we find
\begin{align}
P_{\delta\theta^2} (\omega)
&= \frac{1}{m^2 f_\phi^4}
\int d^3 v_1 d^3 v_2 
\, 
n(\vec v_1) n(\vec v_2) 
\nonumber\\
&\times
\Big[
(2\pi) \delta(\omega - \omega_1 - \omega_2)
+ (2\pi) \delta(\omega + \omega_1 + \omega_2)
\nonumber\\
&
+ (2\pi) \delta(\omega - \omega_1 + \omega_2)
+ (2\pi) \delta(\omega + \omega_1 - \omega_2)
\Big],
\end{align}
where we took the continuum limit in the velocities. 
This is a general expression, which holds for an arbitrary velocity distribution as long as the probability distribution for $\alpha_i$ is given as \eqref{p_a}.

Given a normal velocity distribution \eqref{normal_dist}, we find that the power spectrum of the scalar quadratic operator is
\bea
P_{\delta\theta^2}(f)
= \frac{1}{4} \theta_0^4 \tau_\phi 
\Big[  A(f) + B(f) \Big],
\eea
where $\theta_0 = \sqrt{2\rho_0}/ mf_\phi$, $\tau_\phi = 1/m\sigma^2$ is the coherence time, and 
\begin{align}
A(f) &=
\frac{2\pi \bar v^2}{v_0^2}
\exp\left[ - \frac{\bar v^2 + v_0^2}{\sigma^2}  \right]
I_2 \left( \frac{2v_0 \bar v}{\sigma^2} \right)
\Theta(\bar v^2),
\\
B(f) &=
\frac{2 \sigma}{v_0}
\int_0^\infty dv_c
e^{-\frac{\bar\omega^2}{4v_c^2}}
\left[ e^{-(v_c - \frac{v_0}{\sigma})^2} - e^{-(v_c + \frac{v_0}{\sigma})^2} \right].
\end{align}
Here $I_n(x)$ is the modified Bessel function of the first kind and $\Theta(x)$ is the unit step function. 
For notational simplicity, we have introduced
$$
\bar v^2 = \frac{2\pi f}{m} -2, 
\quad
\textrm{and}
\quad
\bar\omega = \frac{2\pi f}{m\sigma^2}. 
$$
The spectral function $A(f)$ represents the coherent harmonic oscillation at $\omega =2m$.
The spectral function $B(f)$ represents the low-frequency background at $\omega < m \sigma^2$. 
Note that the above PSD is valid for $f >0$.
The low-frequency spectrum $B(f)$ is still valid for $f<0$, but $A(f)$ changes to $A(-f)$.

\begin{figure}[t]
\centering
\includegraphics[width=0.45\textwidth]{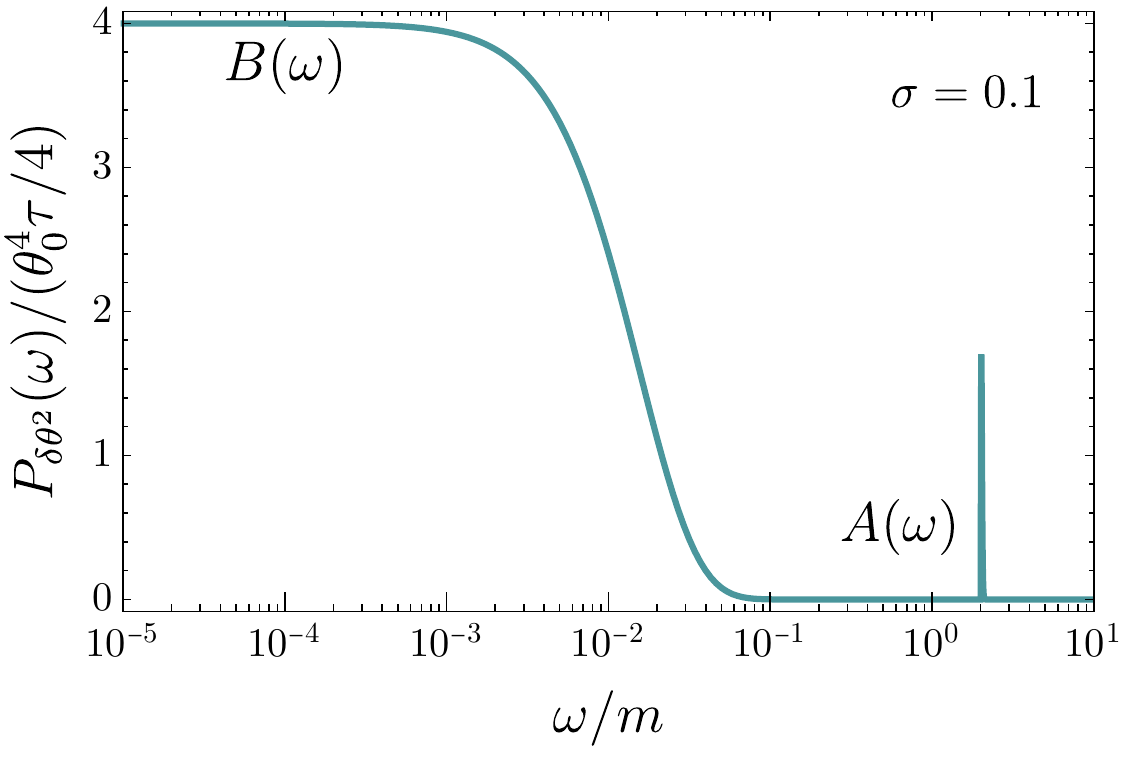}
\caption{The spectrum of the quadratic operator $\delta\theta^2$ in the isotropic limit. 
We choose $\sigma=0.1$ for demonstration. 
The narrow peak at $\omega=2m$ represents the harmonic oscillations, while the plateau at $\omega < m \sigma^2$ gives the low-frequency stochastic background.}
\label{fig:Pdtheta2}
\end{figure}

These expressions are further simplified in the isotropic limit $v_0 \to 0$. 
In this case, we find
\begin{align}
A(f) &= 
\pi \frac{\bar v^4}{\sigma^4} e^{-\bar v^2/\sigma^2}
\Theta(\bar v^2),
\label{A_iso}
\\
B(f) &= 
4\bar\omega K_1( \bar\omega),
\label{B_iso}
\end{align}
where $K_n(x)$ is the modified Bessel function of the second kind and $\Theta(x)$ is the step function. 
Note that both functions, $A$ and $B$, are normalized such that $\tau_\phi \int_0^\infty df \, A (f) = \tau_\phi \int_0^\infty df \, B (f)= 1$. 
The spectrum in this case is shown in Figure~\ref{fig:Pdtheta2}.

\subsection{Allan deviation}\label{app:allan}
Let us consider a single clock comparison test in which the axion causes a signal $s(t)=  K\ \delta\theta^2(t)$. 
If this signal cannot be distinguished from the noise, it still contributes to the total observed variation of the frequencies commonly characterized by the Allan deviation. 
In terms of the fractional frequency shift, the Allan variance over a period $\tau = n \cdot \Delta t$, where $\Delta t$ is the time between measurements, is defined as~\cite{2015RvMP...87..637L}
\begin{align}
    \sigma_s^2(\tau)=\frac{1}{2(M-1)}\sum_{i=1}^{M-1}\left|\langle s(\tau) \rangle_{i+1}-\langle s(\tau)\rangle_{i} \right|^2 ~,
\label{allan_def}
\end{align}
where $\langle s( \tau )\rangle_{i}$ denotes the $i$-th measurement of $s(t)$ over the period $\tau$,
\begin{align}
    \langle s( \tau )\rangle_{i}=\frac{1}{\overline t}\int_{t_i}^{t_i + \tau }dt\ s(t)=K \overline{\delta \theta^2}(t_i)~.
\end{align}
In the second step, we defined $\overline{\delta \theta^2}(t_i)$ as the average value of $\delta \theta^2$ over this period. The ensemble average of the Allan variance then becomes 
\begin{align}
        \langle\sigma_s^2(\tau)\rangle&=\frac{K^2}{2(M-1)}\\
         &\times\sum_{i=1}^{M-1}\left\langle\left|\int df \left(e^{-2\pi i f (t_{i}+ \tau)}-e^{-2\pi i f t_{i}}\right)\widetilde{\overline{\delta \theta^2}}(f)\right|^2 \right\rangle\nonumber\\
         &=2K^2\int_{0}^{\infty} df \sin^2(\pi f \tau )P_{\overline{\delta\theta^2}}(f).
\end{align}
Here the angle bracket denotes an ensemble average.
The Fourier transformation and power spectrum of $\overline{\delta \theta^2}$ are defined analogously to the ones of $\delta \theta^2$. To find the relation between these quantities let us consider the Fourier transformation
\begin{align}
    \widetilde{\overline{\delta \theta^2}}(f)&=\int dt\ e^{2\pi i tf}\left[\frac{1}{\tau}\int_{t}^{t+\tau} dt'\  \delta \theta^2(t')\right]\\
    &=e^{-\pi i ft }\text{sinc}(\pi f \tau)\widetilde{\delta \theta^2}(f)~,
\end{align}
where $\text{sinc}(x)=\sin(x)/x$. The two power spectra are therefore simply related by a factor $\text{sinc}^2(\pi f \tau)$, i.e. $P_{\overline{\delta\theta^2}}(f) = {\rm sinc}^2(\pi f \tau) P_{\delta\theta^2}(f)$.
Using this, we find 
\begin{align}
    \langle\sigma_s^2(\tau)\rangle&= 2K^2\int_{0}^{\infty} df P_{\delta\theta^2}(f)\frac{\sin^4(\pi f \tau)}{(\pi f \tau)^2}~.
\end{align}
From this expression, the Allan deviation caused by the quadratic coupling can be computed using the coupling coefficients that can be found in Tab.~\ref{tab:sensitivity_coeff} and the power spectral density from the last section. 
In particular, in the isotropic limit, we find
\bea
\langle \sigma_s^2(\tau) \rangle
= 2 K^2 \theta_0^4 {\cal I}(\tau/2\tau_\phi) 
\eea
with the integral ${\cal I}(x)$ defined as
\bea
{\cal I}(x) 
= \int_{0}^{\infty} \frac{d\bar\omega}{2\pi}
\, 
\bar\omega K_1(\bar\omega)
\frac{\sin^4 (\bar\omega x) }{(\bar\omega x)^2}. 
\eea
The constraint on $1/f_\phi$ is therefore obtained as
\bea
\frac{1}{f_\phi}
= 
\left[
\frac{m^4 \sigma_{s,\rm obs}^2(\tau)}{8 K^2 \rho_0^2 {\cal I}(\tau/ 2 \tau_\phi)}
\right]^{1/4}
\label{f}
\eea
where $\sigma_{s,\rm obs}(\tau)$ is the experimentally measured Allan deviation with an averaging time $\tau$. 

We obtain the bound shown in Figs.~\ref{fig:summary} and \ref{fig:cross} as a green dashed region, by requiring that the noise caused by the coupling of the axion is below the $1\sigma$ upper bound on the Allan deviation shown in Fig.~1 of \cite{Filzinger:2023zrs} for all given values of $\tau$.

\subsection{Cross-correlation}
Above we computed the correlation between $\delta\theta^2(\omega)$ evaluated at the same spatial position. 
For the cross-correlation of displaced detectors, we must evaluate $\delta\theta^2(\omega)$ at different spatial positions.
In particular, we are interested in
\begin{align}
\langle 
\widetilde{\delta\theta^2}_a(\omega)
\widetilde{\delta\theta^2}^*_b(\omega')
\rangle
= (2\pi) \delta(\omega - \omega')
\frac{1}{2} P^{\rm cross}_{\delta\theta^2}(\omega, \vec L). 
\end{align}
where $\widetilde{\delta\theta^2}_a(\omega) = \widetilde{\delta\theta^2}(\omega,\vec x_a)$ and $\vec L = \vec x_a - \vec x_b$ is the distance between two detectors.
Following the same line of computation, we find
\begin{align}
P^{\rm cross}_{\delta\theta^2} (\omega, \vec L)
&= \frac{1}{m^2 f_\phi^4}
\int d^3 v_1 d^3 v_2 \,
n(\vec v_1) n(\vec v_2) 
\nonumber\\
\times&
\Big[
(2\pi) \delta(\omega - \omega_1 - \omega_2)
e^{+i (\vec k_1 + \vec k_2) \cdot \vec L}
\nonumber\\
&
+ (2\pi) \delta(\omega + \omega_1 + \omega_2)
e^{- i (\vec k_1 + \vec k_2) \cdot \vec L}
\nonumber\\
&
+ (2\pi) \delta(\omega - \omega_1 + \omega_2)
e^{+i (\vec k_1 - \vec k_2) \cdot \vec L}
\nonumber\\
&
+ (2\pi) \delta(\omega + \omega_1 - \omega_2)
e^{-i (\vec k_1 - \vec k_2) \cdot \vec L}
\Big]
\end{align}
Assuming the normal velocity distribution \eqref{normal_dist}, we find
\begin{align}
P^{\rm cross}_{\delta\theta^2}(f, \vec L)
= \frac{1}{4} \theta_0^2 \tau
\Big[ A_{\rm cross}(f,\vec L) 
+ B_{\rm cross}(f, \vec L) 
\Big]
\end{align}
where the two spectral functions are given by
\begin{align}
A_{\rm cross}(f) &=
\frac{2\pi (\bar v/\sigma)^2}{X^2}
\exp\left[ - \frac{\bar v^2 + v_0^2}{\sigma^2}  \right]
I_2 \left( 2X  \frac{\bar v}{\sigma} \right)
\theta(\bar v^2)
\\
B_{\rm cross}(f) &=
2\int_{-\infty}^\infty ds
\frac{e^{-i \bar \omega s}}{(1+s^2)^{3/2}}
\exp\left[
- \frac{( \vec L_\lambda + s \frac{\vec v_0}{\sigma})^2}{1+s^2}
\right]
\end{align}
Here $\vec X = \vec v_0 / \sigma + i \vec L_\lambda$ with $\vec L_\lambda = m \sigma \vec L $ is introduced. 
These expressions reduce to \eqref{A_iso}--\eqref{B_iso} in the isotropic $v_0\to$ and small distance $L \to 0$ limit. 
This expression is valid again for $f>0$.
The low-frequency spectrum $B_{\rm cross}(f,\vec L)$ is valid also for $f<0$, but the other component $A_{\rm cross}(f, \vec L)$ changes for $f<0$ to $A_{\rm cross}(f,\vec L) \to A_{\rm cross}(-f, - \vec L)$.

\bibliography{ref}
\bibliographystyle{apsrev4-1}
\end{document}